\documentclass[aps,prl,twocolumn,showpacs,
amssymb]{revtex4}
\usepackage{graphicx}
\usepackage{dcolumn}
\usepackage{bm}
\usepackage{amsmath,amssymb}

\begin{document}
\newcommand{\MI}{\mathbb{M}}
\newcommand{\RI}{\mathbb{R}}
\newcommand{\HI}{\mathbb{H}}
\newcommand{\CI}{\mathbb{C}}
\newcommand{\TI}{\mathbb{T}}
\newcommand{\EI}{\mathbb{E}}
\def\etal{{\it et al}.} \def\e{{\rm e}} \def\de{\delta}
\def\dd{{\rm d}} \def\ds{\dd s} \def\ep{\epsilon} \def\de{\delta}
\def\goesas{\mathop{\sim}\limits} \def\al{\alpha} \def\vph{\varphi}
\def\Z#1{_{\lower2pt\hbox{$\scriptstyle#1$}}}
\def\X#1{_{\lower2pt\hbox{$\scriptscriptstyle#1$}}}
\def\MM#1{{\cal M}^{#1}} \def\Vm{V(\varphi_0)}
\def\VF{V\Z F} \font\sevenrm=cmr7 \def\ns#1{_{\hbox{\sevenrm #1}}}
\def\Vmin{V\ns{min}} \def\sgn{\,\hbox{sgn}}
\def\coshf{\cosh\left(\chi(u-u\Z0)\right)}
\def\CQG#1{Class.\ Quantum Grav.\ {\bf#1}}
\def\frn#1#2{{\textstyle{#1\over#2}}}

\title{Accelerating cosmologies from compactification with a twist}

\author{Ishwaree P. Neupane$^{1,2,}$
\footnote{E-mail: ishwaree.neupane@canterbury.ac.nz} and David L.
Wiltshire$^{1,}$ \footnote{E-mail:
david.wiltshire@canterbury.ac.nz}} \affiliation{$^{1}$Department
of Physics and Astronomy, University of Canterbury, Private Bag
4800, Christchurch, New Zealand} \affiliation{$^{2}$Central
Department of Physics, Tribhuvan University, Kirtipur, Kathmandu,
Nepal}

\begin{abstract}
It is demonstrated by explicit solutions of the $4+n$--dimensional
vacuum Einstein equations that accelerating cosmologies in the
Einstein conformal frame can be obtained by a time--dependent
compactification of string/M--theory, even in the case that
internal dimensions are Ricci--flat, provided one includes one or more
geometric twists. Such acceleration is transient. When both compact
hyperbolic internal spaces and geometric twists are included,
however, the period of accelerated expansion may be made
arbitrarily large.
\end{abstract}

\pacs{98.80.Cq, 11.25.Mj, 11.25.Yb. \qquad [ {\bf arXiv}: hep-th/0502003 ]}

\maketitle

The observation that the present expansion of the universe is
accelerating \cite{Bennett03a} has proved a challenge to
fundamental theories such as string/M--theory, and its low--energy
supergravity limits. Fundamental scalar fields are abundant in
such higher--dimensional gravity theories, and potentially provide
a natural source for the gravitationally repulsive ``dark energy''
that could explain the present cosmic acceleration, or
alternatively at much higher energy scales the very early period
of cosmic inflation. However, the constraints imposed by
string/M--theory on particular scalar fields, such as the moduli
corresponding to 6 or 7 extra compactified dimensions, are such
that realistic scenarios giving accelerating cosmologies are very
difficult to arrive at.

Indeed for many years, it was assumed that cosmic acceleration was
ruled out for supergravity compactifications on the basis of a
``no go'' theorem \cite{nogo}, which forbids accelerating
cosmologies in the presence of static extra dimensions in
supergravity compactifications, assuming that one wishes to stay
within the realm of classical supergravity rather than resorting
to the addition of ``quantum correction'' terms to the action.
Recently Townsend and Wohlfarth \cite{TW} demonstrated, however,
that it was possible to circumvent the no--go theorem in a
time--dependent supergravity compactification using compact extra
dimensions with negative curvature. Many additional examples
\cite{CHNW03b}--\cite{Bergshoeff} were subsequently found.

The advent of time--dependent compactifications is an important one,
as it has the potential to offer a resolution to the dilemma posed
by the observed cosmic acceleration within a natural theoretical
framework. Nonetheless, a number of substantial problems remain in the models
studied to date. In this Letter we will present new solutions arising
from compactifications with a geometrical twist, which we believe overcome
these problems to give time--dependent supergravity compactifications with
compelling physical features.

Let us turn to the open issues in time--dependent cosmological
compactifications. Firstly, in the Townsend--Wohlfarth (henceforth TW) model
\cite{TW} the no--go theorem was only circumvented by the choice of
negatively curved internal spaces, which are made compact by periodic
identification. While such spaces have particular features, such as
the absence of massless Kaluza--Klein vectors which may have some
phenomenological appeal,
the restriction on the curvature is a severe one, especially given the
view that Ricci--flat internal space compactifications are among the
most natural in string theory.

The second important issue is the question of the length of the period of
accelerated expansion. The TW model and its successors
incorporate scalar fields whose potential energy does not admit a
local minimum with positive vacuum energy. A consequence of this is that
the period of acceleration is transient, corresponding to just a few
e--foldings. Again if one is to apply the model to describe the present
cosmic evolution, then this is not a major difficulty but does place
some constraints on the parameters involved. However, if rather than a
model of the present accelerated cosmological evolution a model of very
early universe inflation via time--dependent compactifications was desired,
then this would be a significant problem.

In this Letter we will present model cosmologies from
time--dependent compactifications which overcome both of the
problems above, while also satisfying a third property of
theoretical naturalness. The limitations with the models studied
to date arise, we believe, from an oversimplification of the
effective potentials that arise from compactifications. In
particular, when a number of scalars associated with different
dimensions are present, their interactions can give rise to
effects which are absent if the extra dimensions form a single
space of constant curvature, as in the TW model. Product space
compactifications have already been considered, sometimes
also in the presence of higher dimensional form--fields with
non--zero fluxes \cite{CHNOW03b}. However, while
higher--dimensional fluxes go a little way towards altering the
form of the compactification potential, for example, by allowing
the presence of a local de Sitter minimum \cite{CHNOW03b}, in the
examples studied to date there is not enough parameter freedom to
allow significant periods of accelerated expansion.

Here we will explicitly consider cosmologies that arise in models
of gravity which correspond to the dimensional reduction to 4 dimensions
of the Einstein equations in 10, 11 or generally $4+n$ dimensions, where
some of the extra dimensions form product spaces {\it with a geometrical
twist}. In particular, consider a $(4+n)$--dimensional metric {\it ansatz}
with $n=p+3$ internal dimensions,
\begin{eqnarray}\label{11dmetric}
&&\ds_{4+n}^2 =\e^{-2\Phi} \ds^2(M_4) +
r\Z1^{\,2}\,\e^{2\phi_1} \ds^2(\MM p_{\ep\X1})\nonumber\\
&&~~~~+\,r\Z2^{\,2}\,\e^{2\phi_2}\, \ds^2(\MM2_{\ep\X2})
+r\Z3^{\,2}\,\e^{2\phi_3}\left(\dd z + \varpi_{\ep\X2}\right)^2\!,
\end{eqnarray}
where $\phi_i=\phi_i(u)$, the parameters $r_i$ define appropriate
curvature radii, $\ds^2(M_4)$ is the metric of the physical
large dimensions in the form
\begin{equation}\label{FRWmetric}
\ds^2 (M_4)= - a^{2\de}(u)\dd u^2 + a^2(u)\dd\Omega^2_{k,3},
\end{equation}
and $\de$ is a constant, the choice of which fixes the nature of
the time coordinate, $u$. The internal space $\MM p$ is a
$p$--dimensional space of constant curvature of sign
$\ep\Z1=0,\pm1$, and the remaining 3 internal dimensions form a
twisted product space, $\MM2\ltimes\MM1$, as follows
\begin{equation} \label{twist}
\vbox{\halign{&$\displaystyle#$\hfil\cr \ds^2(\MM2\Z{+1})&=\dd
x^2+\sin^2x\,\dd y^2,&\varpi\Z{+1}=f\cos x\,\dd y\hfil, \cr
\ds^2(\MM2\Z0)\hfil&=\dd x^2+\dd y^2,\hfil&
\varpi\Z0=\frac{f}{2}(x\dd y- y\dd x), \cr \ds^2(\MM2\Z{-1})&=\dd
x^2+\sinh^2x\,\dd y^2,\quad&\varpi\Z{-1}=f\cosh x\, \dd y,\cr}}
\end{equation}
when $\ep\Z2=+1,0,-1$ respectively, $f$ being the twist parameter.
One chooses the Einstein conformal frame in four dimensions by setting
$\Phi=p\phi_1/2+\phi_2+\phi_3/2$, and so Newton's constant is
time--independent.

To demonstrate the novel features introduced by a geometric twist,
we will begin by presenting a special exact solution in the case
that the physical universe is spatially flat, $k=0$, and both $\MM
p$ and $\MM2$ are Ricci--flat, i.e., $\ep\Z1=0$ and $\ep\Z2=0$,
but with non--zero twist, $f$. This solution to the
$(p+7)$--dimensional vacuum Einstein equations is most readily
written down in the gauge, $\de=3$, used in Ref.~\cite{TW}. In
particular, we find
\begin{eqnarray}\label{soln1}
&&a(u)=a\Z0\,\e^{[(3q+4)c+pc_1]qu/8}\left[\coshf\right]^{q/4},\nonumber\\
&&\phi_2={-\phi_3\over q}=\frn12\ln
\left[\coshf\right]+\frn14(p\,c\Z1+3qc)\,u,\nonumber\\
&&\phi_1=\frac2p\ln\left(\frac{f r\Z3{a\Z0}^3}{\chi{r\Z2}^2}\right)-c\Z1 u,
\end{eqnarray} where
\begin{equation}\label{chi}
\chi \equiv \frn{1}{2}\sqrt{6pq\,c\,c\Z1+3q(3q+4)c^2-p[(4-q)p+8]c\Z1^2/q}.
\end{equation}
and $u\Z0$, $c$ and $c\Z1$ are integration constants. We have ignored
constants which can be absorbed into the $r_i$. The parameter $q$ denotes
the number of twists, and for the metric (\ref{11dmetric})--(\ref{twist})
we take $q=1$ in (\ref{soln1}), (\ref{chi}).

For reality of $\chi$ we require $\al\Z{1-}c<c\Z1<\al\Z{1+}c$, where
$\al\Z{1\pm}\equiv[3\pm\sgn(c)2\sqrt{6(3+7/p)}]/(3p+8)$ when $q=1$. It is
readily seen that the scale factor of the physical universe is monotonic
for all values of $c\Z1$ in this interval. We therefore restrict our
attention to solutions with $c>0$ which correspond to universes which
expand as $u$ increases. Thevalue of $u\Z0$ is merely a gauge choice and
so we set $u\Z0=0$.

In terms of the physical cosmic time, (i.e., proper time of
co--moving observers), $t=\pm \int^u\dd\bar u\, a^3(\bar u)$, both
the early and late time behaviour of the cosmic scale factor is
$a\goesas t^{1/3}$. This is similar to the TW solution with two minor
differences. Firstly, here the $t\to\infty$ limit
corresponds to $u\to\infty$, whereas in the TW solution this
limit corresponds to $u\to0^-$. Secondly, the TW solution decelerates
slightly less quickly at late times, with $a\goesas t^{n/(n+2)}$. Otherwise,
the solution is physically very similar to the TW solution.
In particular, since the acceleration parameter (for $q=1$) is given by
\begin{eqnarray}
{a^5\ddot a}&=&\frn14\chi^2\left(1-\frn32\tanh^2\chi u\right)\nonumber\\
&&\ -\frn1{32}\left(7c+p\,c\Z1\right)\left(4\chi\tanh\chi u +7c+p\,c\Z1
\right)\quad
\end{eqnarray}
where an overdot denotes differentiation w.r.t.\ cosmic time, $t$, it
follows that solutions will exhibit a period of transient acceleration
provided that $\al\Z{2-}c<c\Z1<\al\Z{2+}c$, where $\al\Z{2\pm}\equiv
[1\pm2\sqrt{3(3+7/p)}]/(5p+12)$.
Acceleration occurs on the interval $u_-<u<u_+$, where
\begin{eqnarray}&&
\tanh(\chi u_{\pm})\nonumber\\&&
\equiv\frac{-(7c+p\,c_1)\pm2
\sqrt{7c^2+2p\,c\,c_1-p(5p+12){c_1}^2}}{6\chi}\,.\nonumber
\end{eqnarray}
\begin{figure}[htb]
\vbox{\vskip-10.5pt\centerline{\scalebox{0.75}{\includegraphics{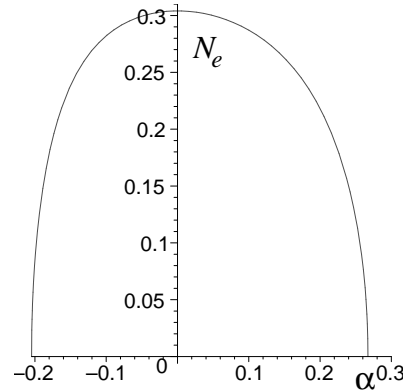}}}
\caption{\label{fig1}%
{\sl The number of e--folds during acceleration epoch\hfil\break $
N_e={q\over8}\ln\left[ \left(1+\tanh\chi u_+\over 1+\tanh\chi
u_-\right)^{\kappa-1} \left(1-\tanh\chi u_-\over 1-\tanh\chi
u_+\right)^{\kappa+1}\right]$ \hfil\break where $\kappa\equiv
[(3q+4)c+pc_1]/(2\chi)$, as a function of the parameter $\al\equiv
c\Z1/c$, for $q=1$, $p=4$, $d=11$.}}}
\end{figure}

The number of e--folds during the period of acceleration,
$N_e\equiv \ln\,a(u_+)-\ln\,a(u_-)$ is plotted in Fig.~\ref{fig1}.
It reaches a maximum $N_e=0.3041$ at $c\Z1=0$ independently of $p$, which
corresponds to the special case in which the internal space $\MM p$ is
static with constant $\phi_1$. (The $p=0$ solution
is formally equivalent to (\ref{soln1})--(\ref{chi})
with $c\Z1=0$ and no $\phi_1$.)
In the general case $\phi_2$ slowly increases --- giving a slow
``decompactification'' as in the TW model \cite{TW} --- while
$\phi_3$ decreases, and $\phi_1$ decreases (increases) if $c\Z1u>0$
($c\Z1 u<0$).

The number of e--folds in the examples is possibly too low to be
consistent with observational bounds, especially when one notes that any
additional matter obeying the strong energy condition would have a
decelerating effect that may decrease the period
of transient acceleration. However, the number of e--folds can be
increased by increasing the dimension of the twisted space.
In particular, consider the case $n=p+2q+1$ when the internal space
is now the product of a $p$--dimensional torus $\MM p$, (i.e., $\ep\Z1=0$),
and the $(2q+1)$--dimensional twisted space
\begin{eqnarray}&&
\ds_{T^{1,\cdots,1}}^2= r_2^2 e^{2\phi_2} \sum_{i=1}^{q}
\left(\dd x_i^2+\dd y_i^2\right) \nonumber\\ &&~~~~~~~~
+ r_3^2 e^{2\phi_3} \left[\dd z+\frac f2\sum_{i=1}^{q}
\left(x_i{\dd y_i}-y_i {\dd x_i}\right)\right]^2\!\!,
\end{eqnarray}
which has topology $(\TI^2\times\cdots\times\TI^2)\ltimes S^1$.
Eqns (\ref{soln1})--(\ref{chi}) with $q$ arbitrary in fact represent
the solution to the $(4+p+2q+1)$--dimensional vacuum Einstein equations
with this more general metric ansatz, where now
$\Phi=p\phi_1/2+q\phi_2+\phi_3/2$. For small values $q>1$ the qualitative
features are the same as the $q=1$ case, with transient acceleration and
a maximum number of e--folds at $c\Z1=0$, independently of $p$. For example,
the maximum number of e--folds for $q=2$, $3$, $4$ are respectively
$N_e=0.3689$, $0.4019$, $0.4221$, a marginal increase.

Next we turn our attention to solutions with a period of
acceleration which is not merely transient.
In addition to a twist, we will take the curvature of the internal space to
be non--zero. To this end, it is convenient to set $\de=0$, so that the
metric~(\ref{FRWmetric}) takes the standard Friedmann--Robertson--Walker
form, and $u$ becomes the cosmic time, $t$. For the metric~(\ref{11dmetric}),
with arbitrary $\ep_i$, the field equations may be
written in the form
\begin{eqnarray}\label{11dEqs}
&&\ddot\Phi+3H\dot{\Phi}-2 K-3\frac{\ddot{a}}{a}=0, \nonumber\\
&&\ddot\Phi+3H\dot{\Phi}-2\left(\frac{\dot{a}}{a}\right)^2
-\frac{\ddot{a}}{a} - \frac{2k}{a^2} = 0,\\
&&\ddot{\phi\Z i}+3 H \dot{\phi}\Z i- \de\Z1 V\Z i-\de\Z2 \VF =
0,\qquad i=1,2,3\nonumber
\end{eqnarray}
where $\de\Z1=2/p$, $\de\Z2=0$ (for $i=1$), $\de\Z1=1$, $\de\Z2=2$
(for $i=2$), $\de\Z1=0$, $\de\Z2=-2$ (for $i=3$),
$H\equiv \dot{a}/a$ is the Hubble parameter, $k$ is the spatial
curvature, and the kinetic and potential terms are respectively
\begin{eqnarray}
&&K=\frn{p(p+2)}4{\dot\phi_1}^2+2{\dot\phi_2}^2+\frn{3}{4}{\dot\phi_3}^2
+p\dot\phi_1\dot\phi_2 + \dot\phi_2\dot\phi_3 +
\frn{p}{2}\dot\phi_3\dot\phi_1, \nonumber\\
&& V=V\Z1+V\Z2+\VF= \Lambda_1\e^{-(p+2)\phi_1-2\phi_2-\phi_3}\nonumber\\
&&~~~~~~~~~~ + \Lambda_2
\e^{-p\phi_1-4\phi_2-\phi_3} + {F}^2
\e^{-p\phi_1-6\phi_2+\phi_3},
\end{eqnarray}
where $\Lambda_1=-p(p-1)\epsilon_1/(2r\Z1^2)$,
$\Lambda_2=-\epsilon_2/r\Z2^2$ and $F\equiv (f/2) (r\Z
3/r\Z2^{\,2})$. Even with $\epsilon_i=0$, the volume modulus field
has a non-zero (positive) potential, and so the cosmological
solutions with $f>0$ circumvent the no-go theorem~\cite{nogo},
while retaining {\it Ricci-flat internal spaces}. Many additional
examples are given in~\cite{Ish05b}.

In terms of alternative canonically normalized scalars, which may
be defined by $\vph_1\equiv-\sqrt{\frn{p+5}{p+4}}\, \phi_3$,\break
$\vph_2\equiv\frn{\sqrt{p(p+4)}}2\left(\phi_1+\frn1{p+4}\phi_3\right)$,
$\vph_3\equiv\frn p2\phi_1+2\phi_2+\frn12\phi_3$, the field
equations are
\begin{eqnarray}
\ddot{\vph}_i+3H\dot{\vph}_i +
\frac{\dd V}{\dd\vph_i}&=&0, \label{waveeq2}\\
\dot{H}+K-ka^{-2}&=&0, \label{commonEqn}
\end{eqnarray}
along with the Friedmann (constraint) equation
\begin{equation}\label{Friedmann1}
H^2= \frn{1}{3}\left(K+V\right)-k a^{-2},
\end{equation}
where $K=\frac{1}{2}\sum_i^3\dot{\vph_i}^2$. The resulting scalar
potential is
\begin{equation}\label{main11dpoten}
V=\Lambda_1\,\e^{-\vph_2/\beta-\vph_3}+\Lambda_2\,
\e^{-2\vph_3}
+F^2\,\e^{-\sqrt{5-\beta^2}\vph_1+\beta\vph_2
-3\vph_3}
\end{equation}
where $\beta\equiv\sqrt{p}/\sqrt{p+4}<1$.
As above we will confine our attention to $k=0$ cosmologies.
Solutions with no twist, $F^2=0$, and compact hyperbolic product
spaces, i.e., $\Lambda_i>0$, have been previously discussed
\cite{CHNOW03b}, and give transient acceleration without a local
minimum in the potential. However, if there is a non--trivial twist
and at least one of $\Lambda_i$ is positive, then it may be
possible to find models with a large number of e-foldings. Indeed,
if we choose $\Lambda_i$ such that the
potential~(\ref{main11dpoten}) is strictly non-negative then it
always has a minimum with respect to a subset of the $\vph_i$ directions.

Even with $k=0$ the general solution to
(\ref{waveeq2})--(\ref{main11dpoten}) is highly non--trivial. To
demonstrate the general physical effects, we will take $\Lambda_2=0$, so
that ${\cal M}^2$ is a $2$-torus, and specialize to the case in which
$d=11$ (or $p=4$) and $\vph_3= b\Z1=$ const. An explicit exact solution
can then be found in terms of a new logarithmic time variable
$\tau$, defined by $\tau=\int^t\dd \bar
t\,\exp\left[-\vph_2(\bar t)/\sqrt{2}\right]$, or
$\vph_2=\sqrt{2}\ln({\dd t}/{\dd\tau}).$ The explicit solution is then
\begin{eqnarray}
&&\left(\ln a\right)^\prime=\frn1{\sqrt{12}}\sqrt{V_0}
\left(\xi+\xi^{-1}\right), \
\vph_2^{\prime}=\frn1{2}\sqrt{V_0}\left(\xi-\xi^{-1}\right),
\nonumber\\
&& \vph_1=\vph_2+\frac{\sqrt{2}b\Z0}{3}, \ \xi\equiv
\frn{\sqrt{6}-1}{\sqrt{5}} \tanh\sqrt{\frac{5V_0}8}(\tau-\tau\Z0),
\end{eqnarray}
where $'\equiv\dd/\dd\tau$, $\tau\Z0$ is a constant,
\begin{equation}
V_0\equiv \Lambda_1\e^{-b\X1}+ F^2 {\rm e}^{-(b\X0+3b\X1)}
=3F^2 {\rm e}^{-(b\X0+3b\X1)}.
\end{equation}
and $b\Z0$ is fixed once $b\Z1$ is chosen.
At late times we find $a\propto t^{2}$, with the corresponding
value of $\omega<-2/3$, where $\omega\equiv (K-V)/(K+V)$. In the
limit $\vph_1\to \vph_2$ we find $\phi_1\to -\phi_2/2$ and $\phi_3\to
-2\phi_2$. Thus two of the extra
dimensions associated with the space ${\cal M}^2$ may grow with time
while other dimensions shrink (or vice versa).

The potential $V$ has a minimum with respect to~$\vph_2$ at
$\vph_{2}^{(0)}=\vph_1+(\sqrt{2}/3) \ln (2\Lambda_1/F^2)+2\sqrt{2}
b\Z0/3$, with
\begin{equation}
V(\vph_0)= \left(\frn{27}{4}\Lambda_1 F^4\e^{-7b\X0}\right)^{1/3}
\,\e^{-\sqrt{2}\vph_1}= V,_{\vph\X2\vph\X2}\!(\vph_0).
\end{equation}
Since the minimum has the curious feature, special to potentials with
$\Lambda_2=0$, that $V,_{\vph\X2\vph\X2}\!(\vph_0)=\Vm$,
particle--like states will have a mass $m^2=\Vm$. If this value were
to be assigned to the vacuum energy at the present epoch, i.e.,
$\Vm\goesas{10}^{-120}$ in Planck units, then both the vacuum energy
and scalar excitations about the vacuum are ultra--light. The degree to
which the model would need to be fine--tuned to achieve such an outcome
remains to be investigated.

Another alternative would be to take $\Vm$ to have the vastly
higher energy scale associated with the epoch of inflation in the
very early universe. Let us assume that $\vph_1$ starts initially
at $\vph_1=\vph_1^{(0)}$, and that $\Lambda_1$ and $F$ both are of
order unity (in Planck units), then when
$\vph_1\goesas\vph_1^{(0)}+47$ the number of e--folds is
$N_e\goesas 65 $. The actual relation between $N_e$ and the shift
in $\vph_1$ can be different, however, depending upon the precise
values of the compactification scale, $\sqrt{\Lambda_1}$, and the
twist parameter, $F$.

A point also worth emphasizing is that, especially in a flat
universe, the de Sitter stage can be transient. If this is the
case in our model, we must allow also $\vph_3$ to roll with
$t$. Different choices of vacua could lead to different asymptotic
expansion. More specifically, for the
potential~(\ref{main11dpoten}), when $k=0$ and $p=4$ the late time
behaviour of the scale factor is characterized by $a(t) \propto
t^\gamma$, where
\begin{equation}\label{valueofp}
\gamma = \left\{\begin{array}{l}
13/19 \quad (\Lambda_2=0,\; \Lambda_1\ne0,\; F^2\ne0)\\
3/5 \qquad (\Lambda_1=0,\; \Lambda_2\ne0,\; F^2\ne0)\\
3/4 \qquad (F^2=0,\;\Lambda_i\ne0)\\
7/9 \qquad (\Lambda_i\ne0,\; F^2\neq 0).
\end{array} \right.
\end{equation}
It is expected the contribution of dust or radiation could modify the above
asymptotic behaviour.

Let us summarize why we believe that exponential potentials
arising from compactifications of supergravity models on a twisted
product space of time-varying volume are a potentially significant source for
dynamical dark energy at the present epoch.
Firstly, Townsend and Wohlfarth \cite{TW} argued that
the no--go theorem for accelerating cosmologies in supergravity
compactifications could only be circumvented by including compact
hyperbolic internal spaces, and would not work in the Ricci--flat
cases. We have shown on the contrary that the no--go theorem
is also circumvented by twisted Ricci--flat spaces by explicit construction
of exact solutions with
the internal space $\TI^p\times(\TI^2\times\cdots\times\TI^2)\ltimes S^1$.
Since Ricci--flat spaces are natural in the string/M--theory context,
this is an important development. Secondly,
we have demonstrated that it is possible to construct a
metastable de Sitter vacuum in the general framework of~\cite{CHNOW03b} by
incorporating one or more geometric twists in the internal space.

As mentioned above, some of the extra dimensions may eventually
grow large, though slowly, without bound and decompactify, a feature
that our transient solutions share with those of the TW model. One
might argue that provided that the relative size of the extra dimensions
and their rate of change remain sufficiently small as to be consistent
with observation over any epoch of physical relevance, then the
eschatological consequences of such ``decompactifications'' are no more
severe than those of a universe whose ordinary three dimensions accelerate
forever. Nonetheless, given that the natural endpoint of the solutions is
a fundamentally higher--dimensional runaway epoch~\cite{Giddings04a}, it
is clear that we would have additional cosmic coincidence constraints in
explaining why the universe appears to have three large spatial dimensions
at the present epoch. However, before arriving at any such conclusions
it is important to include additional matter, such as pressureless
dust. Matter obeying the strong energy condition will certainly affect
the length of any transient acceleration, and may also change the evolution
of the scalar fields.

Finally, we remark that geometric twists add richness to string/M--theory
cosmology, and may potentially lead to the realization of
new cosmological scenarios.

\medskip

\noindent {\bf Acknowledgements}: This work was supported in part
by the Mardsen Fund of the Royal Society of New Zealand. I.P.N.
thanks CERN-TH for hospitality while part of this work was done.



\begin{thebibliography}{99}

\bibitem{Bennett03a} C.L. Bennett \etal,
Astrophys.\ J.\ Suppl.\ {\bf 148}, 1 (2003).

\bibitem{nogo}
G.~W.~Gibbons, 
in {\it Supersymmetry, Supergravity and Related Topics}, 
eds.\ F.~del Aguila, J.~A.~de Azcarraga and L.~E.~Ibanez,
(World Scientific, Singapore, 1985), pp.~123--146;
J.~M.~Maldacena and C.~Nunez,
Int.\ J.\ Mod.\ Phys.\ A {\bf 16}, 822 (2001).

\bibitem{TW}
P.~K.~Townsend and M.N.R.~Wohlfarth,
Phys.\ Rev.\ Lett.\ {\bf 91} (2003) 061302.

\bibitem{CHNW03b} C.-M.~Chen, P.-M.~Ho, I.~P.~Neupane, and J.~E.~Wang,
JHEP {\bf 0307}, 017 (2003).

\bibitem{OhtaPRL03} N.~Ohta,
Phys.\ Rev.\ Lett.\ {\bf 91}, 061303 (2003).

\bibitem{MNR03a} M.~N.~R.~Wohlfarth,
Phys.\ Lett.\ B {\bf 563}, 1 (2003);
S.~Roy,
Phys.\ Lett.\ B {\bf 567}, 322 (2003).

\bibitem{Emparan03a}
R.~Emparan and J.~Garriga,
JHEP {\bf 0305} (2003) 028.

\bibitem{CHNOW03b} C.-M.~Chen, P.-M.~Ho, I.~P.~Neupane, N.~Ohta and
J.~E.~Wang,
JHEP {\bf 0310}, 058 (2003).

\bibitem{IPN03d} I.~P.~Neupane,
\CQG{21}, 4383 (2004);
Mod.\ Phys.\ Lett.\ A {\bf 19}, 1093 (2004);
L.~Jarv, T.~Mohaupt and F.~Saueressig,
JCAP {\bf 0408}, 016 (2004).

\bibitem{Bergshoeff} E.~Bergshoeff, A.~Collinucci, U.~Gran, M.~Nielsen
and D.~Roest,
\CQG{21}, 1947 (2004).

\bibitem{Ish05b}
  I.~P.~Neupane and D.~L.~Wiltshire,
  hep-th/0504135.

\bibitem{Giddings04a} S.~B.~Giddings and R.~C.~Myers,
Phys.\ Rev.\ D {\bf 70}, 046005 (2004).

\end{thebibliography}
\end{document}